\documentclass[aps,prl,showpacs,superscriptaddress,twocolumn,floatfix]{revtex4}
\usepackage{latexsym}
\usepackage{amsfonts}
\usepackage{amssymb}
\usepackage{amsmath}
\usepackage{graphicx}

\begin{document}

\title{Non-uniformity of a planar polarizer as a reason of spin-transfer induced vortex oscillations at zero field}
\author{A. V. Khvalkovskiy}
\altaffiliation{Corresponding author. Electronic address: khvalkov@fpl.gpi.ru}
\affiliation{Unit\'e Mixte de Physique CNRS/Thales and Universit\'e Paris Sud 11, 1 ave A. Fresnel, 91767 Palaiseau, France}
\affiliation{A.M. Prokhorov General Physics Institute, RAS, Vavilova ul. 38, 119991 Moscow, Russia}
\author{J. Grollier}
\affiliation{Unit\'e Mixte de Physique CNRS/Thales and Universit\'e Paris Sud 11, 1 ave A. Fresnel, 91767 Palaiseau, France}
\author{N. Locatelli}
\affiliation{Unit\'e Mixte de Physique CNRS/Thales and Universit\'e Paris Sud 11, 1 ave A. Fresnel, 91767 Palaiseau, France} 
\author{Ya.V. Gorbunov}
\affiliation{A.M. Prokhorov General Physics Institute, RAS, Vavilova ul. 38, 119991 Moscow, Russia}
\author{K. A. Zvezdin}
\affiliation{A.M. Prokhorov General Physics Institute, RAS, Vavilova ul. 38, 119991 Moscow, Russia}
\affiliation{Istituto P.M. s.r.l., via Cernaia 24, 10122 Torino, Italy}
\author{V. Cros}
\affiliation{Unit\'e Mixte de Physique CNRS/Thales and Universit\'e Paris Sud 11, 1 ave A. Fresnel, 91767 Palaiseau, France}

\begin{abstract}
We discuss a possible mechanism of the spin-transfer induced oscillations of a vortex in the free layer of spin-valve nanostructures, in which the polarizer layer has a planar magnetization. We demonstrate that if such planar polarizer is essentially non-uniform, steady gyrotropic vortex motion with large amplitude can be excited. The best excitation efficiency is obtained for a circular magnetization distribution in the polarizer. In this configuration, the conditions for the onset of the oscillations depend on the vortex chirality but not on the direction of its core.\end{abstract}

\maketitle

Sub-GHz dynamics of magnetic vortices induced by the spin transfer effect observed recently in nanopillars and nanocontacts \cite{Pribiag,Pufall,Mistral,Julich,Theodoropoulou,Pribiag09} have raised a strong interest. The associated microwave emissions in such Spin Transfer Vortex Oscillators (STVOs) occur without any external magnetic field and at low current densities, together with large powers (in the nW range) and narrow linewidths ($<$ 1 MHz) comparatively to single-domain spin transfer nanooscillators. In arrays of nanocontacts, a coherent motion of coupled vortex dynamics generated by the spin transfer, resulting in a significant improvement of the quality factor of the devices has been recently demonstrated \cite{Ruotolo}. This makes STVOs of considerable practical interest for new set of applications in microwave technologies or magnetic memories.

STVO consists of at least two magnetic layers separated by a non-magnetic spacer. One of the magnetic layers (the free layer) has a vortex that can be excited by the spin transfer, while the second magnetic layer is used as a spin polarizer for the current. So far, the theoretical analysis of the spin transfer vortex dynamics has only considered the approximation of a fixed uniformly magnetized polarizer \cite{IvanovSTT, Liu, Mistral, KhvalkovskiyPRB, GuslienkoSTT}. For such polarizers, only the component of the spin polarization that is perpendicular to the plane can induce steady vortex precession \cite{Dussaux}. However, in many recent experiments spin transfer driven vortex oscillations have been detected at zero or in-plane bias magnetic field, in nanopillar or nanocontact STVOs for which the magnetization of the polarizer naturally lies in the plane \cite{Pribiag,Ruotolo,Julich,Theodoropoulou,Pribiag09}. The onset of a small perpendicular component of the spin polarization due to the magnetization dynamics in the polarizer can be assumed, but such contribution can not be sufficient to account for the large amplitude vortex excitations. 

In our present work we consider another mechanism for the vortex excitation, which is specifically related to the STVO having planar magnetization distribution within the polarizer. First we present an analytical model for the vortex dynamics in a circular spin-valve nanopillar. The free magnetic layer of the spin valve is in a centered vortex state. The second magnetic layer (the polarizer) is magnetized in the layer plane, which leads to an in-plane spin polarization. To clarify our analysis, we disregard the stray magnetic field emitted by the polarizer and assume that it is fixed. The current flow is assumed to be uniform through the pillar diameter. The spin transfer torque \cite{Slonc} is calculated using $(\sigma J / M_s) \mathbf{M} \times (\mathbf{M} \times \mathbf{p})$, where $J$ is the current density, $M_s$ is the magnetization of saturation, $\mathbf{M}$ is the magnetization vector, $\mathbf{p}$ is the spin polarization vector, and $\sigma$ represents the efficiency of the spin transfer: $\sigma = \hbar \nu / (2 \left|e\right| L M_s)$, $\nu$ is the spin polarization of the current, e is the electron charge, $L$ the layer thickness. 

We have recently suggested to use the energy dissipation approach to derive the generalized Thiele equation for the spin current-induced vortex core motion \cite{KhvalkovskiyPRB}:
\begin{equation}\label{ThieleEquation}
\mathbf{G}\times\frac{d\mathbf{X}}{dt} - \frac{\partial W}{\partial \mathbf{X}} - D \frac{d\mathbf{X}}{dt} + \mathbf{F}_{ST}=0. 
\end{equation}
Here $\mathbf{X}$ is the vortex core position, the gyrovector is given by $\mathbf{G}$ = $- G \mathbf{e_{z}}$, with $G$ = $2\pi M_s L / \gamma$ and $W(\mathbf{X})$ is the potential energy of the moving vortex \cite{GuslienkoJAP2002}. The damping constant $D$ is given by $D$ = $\alpha \eta G$, where the factor $\eta$ is of the order of unity \cite{GuslienkoD}. The last term $\mathbf{F}_{ST}$ is the spin transfer force. 

We derive the terms of Eq. \ref{ThieleEquation} using the standard \textit{two-vortices ansatz} (TVA) for the in-plane magnetization components of the moving vortex \cite{GuslienkoJAP2002}. The out-of-plane magnetization component $M_s \cos \theta$ can be approximated by a bell-shaped ansatz suggested by Usov et al. \cite{Usov}:
\begin{equation}\label{UsovAnsatz}
 \theta =  \left\{ 
				\begin{array}{ll}
				2 P \tan^{-1} \left( \rho / b \right) & (\rho < b )\\
				P \pi /2 & (\rho \geq  b)
				\end{array}
\right. 
\end{equation}
where $b$ is the core radius and $P$ is the polarity of the vortex core ($P= +1$ if the polarity is along the $z$ axis and $P = -1$ if it is opposite). We consider in these calculations a steady-state circular motion of the vortex core:
\begin{equation}\label{GyroMotion}
\dot{\mathbf{X}} = \omega \mathbf{e}_{z} \times \mathbf{X},\ b << a << R.
\end{equation}
where $a$ is the orbit radius, $\omega$ is the gyration frequency. The vortex energy $W(\mathbf{X})$ has two main contributions: the magnetostatic energy, which originates from the volume magnetic charges arising from a shifted vortex \cite{GuslienkoJAP2002}, and the contribution of the Oersted field \cite{KhvalkovskiyPRB,YSChoi2}. The last two terms of Eq. (\ref{ThieleEquation}) can be calculated using the energy dissipation function $\dot{W}= \int \dot{w}\left(\mathbf{r}\right) dV$, $w$ is the energy density at point $\mathbf{r}$. We use $\dot{w}$ = $\left(\delta E / \delta \theta \right) \dot{\theta} + \left(\delta E / \delta \varphi \right)\dot{\varphi}$ \cite{IvanovSTT}, where $\theta$ and $\varphi$ are respectively the polar and azimuth angles of the magnetization vector $\mathbf{M}$. $\delta E / \delta \theta $ and $\delta E / \delta \varphi $ are taken from the LLG equation. This gives us the current-dependent contribution to the energy dissipation density $\dot{w}_{ST}$ for the planar polarizer: 
\begin{eqnarray}\label{w_ST}
\dot{w}_{ST} = M_s \sigma J \left[ \left( p_x \sin \varphi - p_y \cos \varphi \right)\dot{\theta} \right. \nonumber \\
\left. + \sin \theta \cos \theta \left( p_x \cos \varphi + p_y \sin \varphi \right)\dot{\varphi} \right]
\end{eqnarray}
where p$_x$ and p$_y$ are the $x$- and $y$-components of the local spin polarization $\mathbf{p}$. The right-hand side of Eq. (\ref{w_ST}) vanishes outside the core, from which we find that for the planar polarizer the spin torque excites only the vortex core. The spin-transfer force is given by $\mathbf{F}_{ST}$ $=$ $\partial \left( \int \dot{w}_{ST}dV \right) / \partial \dot{\mathbf{X}}$. Using Eq. (\ref{UsovAnsatz}) and (\ref{GyroMotion}), we find:
\begin{equation}\label{F_ST}
\mathbf{F}_{ST} = \pi M_s L b P \sigma J \left( \mathbf{p} \left( \mathbf{X} \right) \cdot \mathbf{e_{\chi}} \right)\mathbf{e_{\chi}},
\end{equation}
in which the terms proportional to $b^2$ have been neglected. The damping force $\mathbf{F}_{damp}$ $\equiv$ $- D \dot{\mathbf{X}}$ can be obtained by treating similarly the contribution to $\dot{W}$ proportional to the Gilbert damping $\alpha$ \cite{KhvalkovskiyPRB}:
\begin{equation}\label{F_damp}
\mathbf{F}_{damp} = - \eta G P a \omega \mathbf{e_{\chi}}.
\end{equation}
The vortex energy gain, given by the dot product $\left( \mathbf{F}_{ST}+\mathbf{F}_{damp} \right) \cdot \dot{\mathbf{X}}$, should average to zero in each cycle of the steady core gyration. This leads to the following general condition for the onset of the steady vortex precession for arbitrary magnetization distribution $\mathbf{p} \left(a,\chi\right)$ in a planar polarizer: 
\begin{equation}\label{GeneralCond}
\frac{2 \pi \alpha \eta \omega a}{ \ln 2 \gamma \sigma J b} = \int^{\chi= 2 \pi}_{\chi = 0} \left( \mathbf{p} \left(a,\chi\right) \cdot \mathbf{e_{\chi}} \right) d \chi. 
\end{equation} 
If the magnetization in the planar polarizer is uniform, i.e. $\mathbf{p}\left(\chi\right)$ = $const$, the spin transfer torque contributes positively  to the energy gain for one semi-cycle of the vortex motion, $ \left( \mathbf{p} \cdot \mathbf{e_{\chi}} \right) > 0$, but it contributes negatively and with the same amplitude for the other semi-cycle, $ \left( \mathbf{p} \cdot \mathbf{e_{-\chi}} \right) < 0$. Therefore a uniform planar polarizer should not excite the steady vortex motion.

Nevertheless the vortex motion can be excited if $\mathbf{p}\left(\chi\right)$ is a non-trivial function. We conclude from Eq. (\ref{GeneralCond}) that the planar polarizer is the most efficient when $\mathbf{p}$ has a circular distribution, that is $\left(\mathbf{p} \cdot \mathbf{e}_{\chi}\right) = \pm 1$ for each $\chi$, corresponding to a centered vortex in the polarizer layer \cite{Pribiag09}.  For such a circular planar polarizer, we find from Eq. (\ref{GeneralCond}) an expression for the radius of the vortex core orbit $a$ as a function of the current density $J$: 
\begin{equation}\label{AvsJ}
a = C_v C_{pol} \frac{\gamma b \sigma \ln 2}{\alpha \eta \omega} J.
\end{equation} 
where $C_v$ and $C_{pol}$ are the chiralities of the vortex respectively in the free and polarizer layer. From Eq. (\ref{AvsJ}), we conclude that, at a given current sign, the onset of the oscillations ($a > 0$), is not sensitive to the vortex core polarity $P$ but depends on the chirality of the vortex $C_v$. This feature is different from the conditions for spin transfer vortex excitations by a perpendicular polarizer as discussed below. 

We now compare our analytical results to numerical micromagnetic simulations, which have been performed for a nanopillar of 200 nm in diameter with a free NiFe layer of a thickness of 15 nm. We use the following magnetic parameters: $M_s$ = 800 emu/cm$^3$, A = 1.3 $\times$ 10$^{-6}$ erg/cm and $\alpha$ = 0.01 (values for NiFe) and a mesh with the cell size 2 $\times$ 2 $\times$ 5 nm$^3$. The spin current polarization is taken to be $\nu$ = 0.3. The micromagnetic simulations are performed by numerical integration of the LLG equation using our micromagnetic code SpinPM based on the forth order Runge-Kutta method with an adaptive time-step control for the time integration. 
\begin{figure}[h]
   \centering
    \includegraphics[keepaspectratio=1,width=8.5 cm]{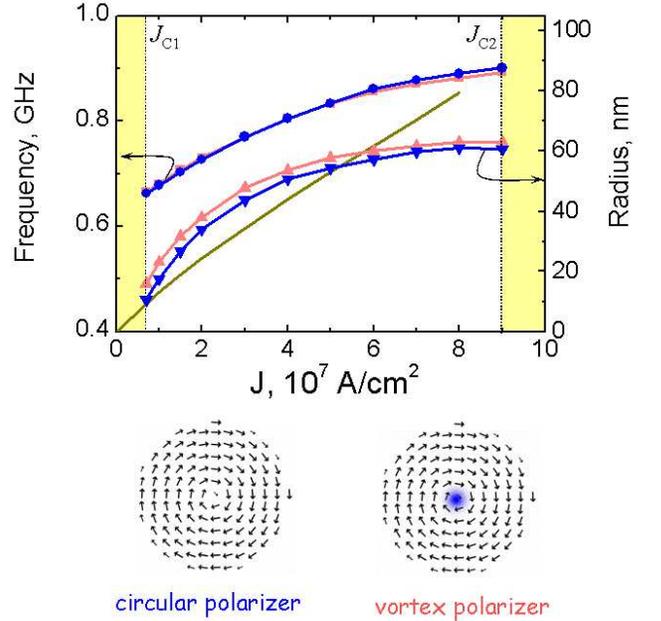}
     \caption{ (color online) Numerical result for spin current-induced vortex gyration for the vortex polarizers, shown at the bottom (colorscale shows $z$-component of the polarization). Graph: frequency $f$ (left scale; circles: circular polarizer, squares: vortex polarizer) and radius of the vortex core orbit $a$ (right scale; down triangles: circular polarizer, up triangles: vortex polarizer) as function of the current density $J$. Solid line shows prediction for $a(J)$ by Eq. \ref{AvsJ}.}
\label{Fig1}
\end{figure}

We assume that the chirality of the vortex in the free layer $C_v$ is set by the symmetry of the Oersted field \cite{note1}. The polarizer layer has an artificially designed perfectly circular magnetization distribution, with $\left| \left( \mathbf{p}, \mathbf{e}_{\chi}\right)\right|= 1$ for each point; thus it lies in-plane even in the disc center. We refer to such an idealized planar configuration as a \textit{circular polarizer}. Its chirality is $C_{pol}$. The positive current is defined as a flow of electrons from the free layer to the polarizer. 

For $C_v C_{pol} = 1$, we see that the vortex motion is excited at positive currents, and for $C_v C_{pol} = -1$ it is excited at negative currents. We also find that results of the simulation are identical for both polarities of the vortex core. As shown in Fig. (\ref{Fig1}), a steady circular motion is observed for current densities larger than $J_{C1}$ $=$5 $\times$ 10$^6$ A/cm$^2$ and smaller than $J_{C2}$ $=$ 1.0 $\times$ 10$^8$ A/cm$^2$. For $J_{C1}< J < J_{C2}$, the frequency increases with $J$ from 0.66 GHz up to 0.90 GHz. The radius of the vortex gyration also increases with the current, with the maximum value of about 60 nm. For currents densities larger than $J_{C2}$, the vortex polarity is periodically switching. At each switching event, the vortex core starts to move in the opposite direction \cite{Kim07}. However since the spin current provides the excitation for both polarities on equal basis, the core is again accelerated by the spin torque until its velocity reaches the critical value required for the reversal \cite{GuslienkoPRL08,KSLee}. 

These numerical results are in very good agreement with the analytical conclusions. First, they confirm the possibility to excite a spin-transfer induced vortex motion for a purely planar polarizer. We find that the current sign for which the motion can be excited depends on the product $C_v C_{pol}$ but not on the core polarity $P$, in agreement to the predictions of Eq. (\ref{AvsJ}). In Fig. \ref{Fig1}, we plot as a solid line the prediction for the orbit radius calculated from Eq. (\ref{AvsJ}) which is in reasonable agreement with the numerical result \cite{note2}. Some quantitative difference between the analytical and the numerical calculations can be ascribed to the considerable deviation of the magnetization distribution from the TVA for $a > b$, as analyses of the magnetization distributions reveal.

Both the analytical and numerical results are qualitatively different from what has been found for the case of the vortex excitation by a uniform out-of-plane polarization. In that case, the spin-transfer force originates from the out-of-the-core region of the vortex (in contrast to the case of the planar polarizer, for which the force originates from the core). It is given by $\mathbf{F}_{ST}^{perp}$ = $\pi M_s p_z L \sigma J a \mathbf{e_{\chi}}$ \cite{KhvalkovskiyPRB, GuslienkoSTT}, where $p_z$ is the out-of-plane spin polarization. It does not depend on the vortex polarity $P$. Therefore, at a given current sign, it can excite the vortex motion (i.e. it can be oppositely directed to the $P$-dependent damping force $\mathbf{F}_{damp}$) for only one vortex polarity. In the steady oscillation regime, the radius of the vortex orbit $a$ is inversely proportional to small non-linear terms in $\mathbf{F}_{ST}^{perp}$ and $\mathbf{F}_{damp}$ \cite{IvanovSTT,GuslienkoSTT}. Due to this reason $a$ can increase very rapidly with the current for $J > J_{C1}$ as we found recently  \cite{KhvalkovskiyPRB}. In contrast to it, for the planar polarizer for $J > J_{C1}$, $a(J)$ is proportional to the principle values of the forces , see Eq. (\ref{AvsJ}); this is the reason of the considerably slower dependence of $a$ on $J$ found in the simulations. 

The closest experimental situation to the idealized \textit{circular polarizer} is the polarizer in the vortex state \cite{Pribiag09}. For such polarizer, an additional contribution to the spin transfer term might arise from the out-of-plane core region. We performed micromagnetic simulations for such configuration which we refer to as a \textit{vortex polarizer}. We find, see Fig.\ref{Fig1}, that the oscillation frequency is practically identical to the \textit{circular polarizer} case, while the radius of the trajectory is only slightly shifted towards higher values. This result allows us to rule out the contribution of the vortex core in the planar polarizing layer as the major source of polarization for the spin transfer force.

As a conclusion, we demonstrate both analytically and by micromagnetic simulations that a non-uniform planar polarizer can induce spin-transfer vortex oscillations. In the case of an ideal circular polarizer, we have derived the conditions for sustained vortex precession in the free layer as a function of the current signs and the respective vortex chiralities. Important conclusion is that the condition for spin transfer precession does not depend on the core polarity. In addition, at large currents, a multiple back and forth core switching during the motion is predicted. 

This work is supported by the EU project MASTER (No. NMPFP7 212257), the French ANR project VOICE (PNANO-09-P231-36) and RFBR Grant (No. 10-02-01162).


\begin{thebibliography}{32}
\bibitem{note1}
To check different signs of the current without changing the chirality of the Oersted field, we consider that the free layer can be above as well as below the polarizer layer in the stack.
\bibitem{note2}
We obtain the core radius $b$ by fitting the simulated static vortex profile with the Ansatz (\ref{UsovAnsatz}), which gives $b$ = 16 nm and the parameter $\eta$ = 1.46 is found using the numerical procedure described in Ref. \cite{KhvalkovskiyPRB}. The first critical current can be captured by the analytical model by taking into account the higher order terms in $W(a)$ and $\dot{W}(a)$ for $a \approx b$, an issue that is going beyond the scope of the present work.
\end{thebibliography}
\end{document}